\documentclass[12pt, a4paper]{article}
\usepackage{a4wide,amssymb,amsmath,amscd}

\begin{document}

\topmargin -2pt

\headheight 0pt

\begin{flushright}
{\tt KIAS-P08002}
\end{flushright}

\vspace{5mm}

\begin{center}
{\Large \bf Kerr-Newman-de Sitter Solution on DGP Brane}\\

\vspace{10mm}

{\sc Daeho Lee}${}^{  \dag,  }$\footnote{dhlee@sju.ac.kr}, {\sc Ee Chang-Young}${}^{ \dag,
 }$\footnote{cylee@sejong.ac.kr}, and {\sc Myungseok Yoon}${}^{  \ddag,  }$\footnote{younms@sogang.ac.kr}
\\

\vspace{1mm}

 ${}^{\dag}${\it Department of Physics, Sejong University, Seoul 143-747, Korea}\\

${}^{\ddag}${\it Center for Quantum Spacetime, Sogang
    University, Seoul 121-742, Korea}\\

\vspace{10mm}

{\bf ABSTRACT} \\
\end{center}


\noindent

We find an exact solution of Kerr-Newman-de Sitter type on the braneworld(4D) of the DGP model.
When a constant 4D Ricci scalar is assumed, only zero(flat) and a positive(de-Sitter) values
satisfy the Hamiltonian constraint equation coming from the extra dimension.
With a $Z_2$-symmetry across the brane and a stationary and axisymmetric metric ansatz
on the brane, we solve the constraint equation exactly in the Kerr-Schild form with de-Sitter background.
In the de-Sitter background this Kerr-Schild solution is well behaved under Boyer-Lindquist transformation:
the constraint equation is preserved under the transformation and so is the solution.
In the non-rotating limit we show that this Kerr-Newman-de Sitter solution
has the characteristic of accelerated expansion of the braneworld universe.
\\

\vfill

\noindent
PACS: 04.40.Nr, 04.50.+h, 04.70.-s\\

\thispagestyle{empty}

\newpage




%

\section{Introduction}

Recent astronomical observations of Type Ia supernovae suggested that our universe is
expanding at an accelerating pace \cite{sauv}.
In parallel with this development, the idea
that our universe may be a brane embedded in some higher dimensional space
also became quite popular \cite{rs99}.
One of the models along this line proposed by Dvali, Gabadadze and Porrati
(DGP) \cite{dgp} is known to contain a branch of solutions exhibiting
self-accelerated expansion of the universe \cite{df}. In this model, the
acceleration takes place without having the cosmological
constant on the brane.
Although there have been many works and much interest in the DGP model,
its exact solutions are not much known so far: It's approximate Schwarzschild
solutions had been obtained in \cite{gru,po,ls,ms,nr}, and an exact Schwarzschild solution
on the brane was obtained in \cite{gi:prd}.
Recently, an exact solution for charged
black holes on the brane was obtained in \cite{el},
and we worked out a particular solution for charged rotating black holes on the brane
in \cite{ley}.
However, in our previous work the metric in our modified Boyer-Lindquist coordinates
could not be fully diagonalized: it contains $t$-$\theta$ and $\theta$-$\phi$ cross terms
besides the usual $t$-$\phi$ cross term for rotating solutions.
This causes a difficulty in interpretation of the structure of the horizon.

On the other hand in \cite{ley} we also noticed that the particular solution we obtained
shows the characteristic of Kerr-Newman-de Sitter black hole in general relativity.
Based on this observation, we further investigate along this direction
and in this paper we obtain an exact solution of Kerr-Newman-de-Sitter type with the metric
in conventional diagonalization with $t$-$\phi$ cross term only.

In doing this, we first notice that only two solutions,
flat and de-Sitter type geometries, are allowed by the constraint equation
if constant curvature scalar is assumed.
Then we solve the constraint equation in the Kerr-Schild form \cite{ks} with
de-Sitter background and find an exact solution.
This works well with the Boyer-Lindquist type coordinates \cite{glpp} in the
conventional sense in which only $t$-$\phi$ cross term appears in the metric.
This solution also matches well for a cosmological solution of accelerated expansion:
 Following the work of \cite{lss}, we show that our de-Sitter type solution
exhibits the characteristic of accelerated expansion in the non-rotating limit.

This paper is organized as follows. In section 2, we set the
action and equations of motion of the DGP model following the
approach of \cite{ag:cqg}. In section 3, we obtain a Kerr-Newman type exact solution
on the brane with de-Sitter background in the Kerr-Schild form
and transform it to the Boyer-Lindquist coordinates.
In section 4, we show that our de-Sitter type solution  matches
a cosmological solution with accelerated expansion in the non-rotating limit.
We conclude in section 5.
\\

\section{Field equations on the brane}

The DGP gravitational action in the presence of sources takes the
form \cite{dgp}
\begin{eqnarray}
\label{action}
S = M_{*}^{3}\int d^5 x \sqrt{-g}~^{(5)}R + \int d^4 x \sqrt{-h}
\left(M_{P}^{2}R+L_{matter}\right),
\end{eqnarray}
where $R$ and $^{(5)}R$ are the 4D and 5D Ricci scalars,
respectively and $L_{matter}$ is the Lagrangian of the matter
fields trapped on the brane. Here, the $(4+1)$ coordinates are
$x^{A}=(x^{\mu},y(= x^5))$, $\mu=0,1,2,3$, and $g$ is the determinant
of the five-dimensional metric $g_{AB}$, while $h$ is the determinant
of the four-dimensional metric $h_{\mu\nu}=g_{\mu\nu}(x^{\mu},y=0)$.
A cross-over scale is defined by $r_{c}=
m_{c}^{-1}=M_{P}^{2}/2M_{*}^{3}$.
There is a boundary(a brane) at $y=0$ and $Z_{2}$ symmetry across
the boundary is assumed.
 The field equations derived from the action (\ref{action}) have
the form
\begin{equation}
\label{eom:munu}
^{(5)}G_{AB}= {}^{(5)}R_{AB}-\frac{1}{2}g_{AB}~^{(5)}R=
\kappa_{5}^{2}\sqrt{\frac{h}{g}}\left(X_{AB}+T_{AB}\right)
\delta(y),
\end{equation}
where $\kappa_{4}^{2}=M_{P}^{-2}$ and $\kappa_{5}^{2}=M_{*}^{-3}$,
while $X_{AB}=-\delta_{A}^{\mu}\delta_{B}^{\nu}G_{\mu\nu}/\kappa_{4}^{2}$
and $T_{AB}= \delta_{A}^{\mu}\delta_{B}^{\nu}T_{\mu\nu}$ is the
energy-momentum tensor in the braneworld.

 Now, we consider the metric in the following form \cite{ag:cqg,dgl:jcap},
\begin{equation}
\label{metric}
ds^2=g_{AB}dx^{A}dx^{B}=g_{\mu\nu}(x,y)dx^{\mu}dx^{\nu}+2N_{\mu}
dx^{\mu}dy+(N^2+g_{\mu\nu}N^{\mu}N^{\nu})dy^2.
\end{equation}
The $(\mu 5)$, $(55)$ components of the field equations
(\ref{eom:munu}) are called as the momentum and Hamiltonian
constraint equations, respectively, and are given by
\cite{ag:cqg,dgl:jcap}
\begin{equation}
\label{eom:5i}
\nabla_{\nu}K^{\nu}_{~\mu}-\nabla_{\mu}K=0,
\end{equation}
\begin{equation}
\label{eom:55}
R-K^{2}+K_{\mu\nu}K^{\mu\nu}=0,
\end{equation}
where $K_{\mu\nu}$ is the extrinsic curvature tensor defined
by
\begin{equation}
\label{kmunu}
K_{\mu\nu}=\frac{1}{2N}(\partial_{y} g_{\mu\nu}-\nabla_{\mu}
N_{\nu}-\nabla_{\nu}N_{\mu}),
\end{equation}
and $\nabla_{\mu}$ is the covariant derivative operator
associated with the metric $g_{\mu\nu}$.

Integrating both sides of the field equation (\ref{eom:munu})
along the $y$ direction and taking the limit of $y=0$ on the
both sides of the brane we get the Israel's junction
condition \cite{israel} on the $Z_2$ symmetric brane as follows \cite{el}:
\begin{equation}
\label{jc}
G_{\mu\nu}=\kappa_{4}^2 T_{\mu\nu}+m_{c}(K_{\mu\nu}
-h_{\mu\nu}K).
\end{equation}

In this paper, we take the electro-magnetic field as the
matter source on the brane.
Plug the junction condition (\ref{jc}) back into the constraint equations (\ref{eom:5i}) and
(\ref{eom:55}), we find that the momentum constraint equation
is satisfied identically, while the Hamiltonian constraint
equation is given by
\begin{eqnarray}
\label{Hamiltonc}
 R_{\mu\nu}R^{\mu\nu}-\frac{1}{3}R^{2}+ m_c^2 R+
 \kappa_{4}^{4}T_{\mu\nu}T^{\mu\nu}
 -2\kappa_{4}^2 R_{\mu\nu}T^{\mu\nu}=0,
\end{eqnarray}
where we used $T=T^{\mu}_{~\mu}=0$.
Here we would like to note that when the spacetime has
constant curvature scalar only two values are allowed,
$R=0$(flat) and $R=12m_c^2$(de-Sitter).
This can be seen as follows:
 For $R=4\alpha$ with some constant $\alpha$, one can set
 $R_{\mu\nu}:=\alpha h_{\mu\nu} + \kappa_{4}^{2}T_{\mu\nu}$.
Plug this relation back to the constraint equation (\ref{Hamiltonc}),
one gets $\alpha=0$ or $\alpha=3 m_c^2$.

Finally, combining the Einstein equations in the
bulk($y\neq 0$),
\begin{equation}
^{(5)}G_{AB}= ^{(5)}R_{AB}-\frac{1}{2}g_{AB}~^{(5)}R=0,
\end{equation}
with (\ref{jc}) we arrive at the gravitational field
equations on the brane \cite{el}
\begin{eqnarray}
\label{grave}
G_{\mu\nu} &=&-E_{\mu\nu}-\frac{\kappa_{4}^{4}}{m_{c}^2}
(T^{\rho}_{~\mu}T_{\rho\nu}-\frac{1}{2}h_{\mu\nu}
T_{\rho\sigma}T^{\rho\sigma})
      \nonumber \\
     && -\frac{1}{m_{c}^2}(R^{\rho}_{~\mu}R_{\rho\nu}
-\frac{2}{3}RR_{\mu\nu}+\frac{1}{4}h_{\mu\nu}R^2
-\frac{1}{2}h_{\mu\nu}R_{\rho\sigma}R^{\rho\sigma})
  \nonumber \\
 &&+\frac{\kappa_{4}^2}{m_{c}^2}(R^{\rho}_{~\mu}T_{\rho\nu}
+T^{\rho}_{~\mu}R_{\rho\nu}-\frac{2}{3}RT_{\mu\nu}
-h_{\mu\nu}R_{\rho\sigma}T^{\rho\sigma}),
\end{eqnarray}
where $E_{\mu\nu}$ is the traceless ``electric part'' of the
5-dimensional Weyl tensor $^{(5)}\!C_{ABCD}$ \cite{sms:prd}
and $m_{c}^{-1}=\kappa_{5}^2 /2\kappa_{4}^2$.
In what follows we shall set $\kappa_{4}^2=8\pi$.
 Here, we would like to check   the existence of the gravitational effect from the extra dimension.
 Use the fact that the Hamiltonian constraint (\ref{Hamiltonc})  is satisfied by the relation
\begin{eqnarray}
\label{chamilton}
R_{\mu\nu}=\alpha h_{\mu\nu}+\kappa_{4}^2 T_{\mu\nu},
\end{eqnarray}
for $\alpha=0$ or $3m_c^2$, then the gravitational field equations (\ref{grave}) on the brane reduce to the simple form
\begin{eqnarray}
\label{cgrave}
R_{\mu\nu}=\alpha h_{\mu\nu}-E_{\mu\nu}.
\end{eqnarray}
Comparing (\ref{cgrave}) with (\ref{chamilton}) we can identify the projected Weyl tensor $E_{\mu\nu}$ with the energy-momentum tensor $T_{\mu\nu}$ on the brane as
\begin{eqnarray}
\label{reweyl}
E_{\mu\nu}=-\kappa_{4}^2~T_{\mu\nu}.
\end{eqnarray}
This shows that for the two special cases of $R=0$ and $R=12m_c^2$
 the tensor $E_{\mu\nu}$ has only contribution from the energy-momentum tensor on the brane.
Therefore our field equations on the brane, (\ref{grave}) contains no effect due to the extra dimension
when the spacetime has constant curvature scalar.
This is different from the RS model case  \cite{ag:prd}, where the so-called tidal charge effect due to
the extra dimension appears.
\\

\section{Kerr-Newman solutions on the brane}

In this section, we consider the case in which the brane contains a
Maxwell field with an electric charge. In general, rotating black
holes on the brane can carry electric charges. We assume that
the Maxwell field on the brane is described by a solution of
source-free Maxwell equations.
In that case, the trace of the energy-momentum tensor for the Maxwell
field on the brane vanishes, and we have to solve equation (\ref{Hamiltonc})
and the Maxwell equations:
\begin{equation}
\label{fMaxwell}
g^{\mu\nu}D_{\mu}F_{\nu\sigma}=0,
\end{equation}
\begin{equation}
\label{sMaxwell} D_{[\mu}F_{\nu\sigma]}=0,
\end{equation}
where $D_{\mu}$ is the covariant derivative operator associated
with the brane metric $h_{\mu\nu}$.
However, since (\ref{sMaxwell}) is satisfied identically once the field strength
is expressed in terms of the potential one-form, we only have to
solve equations (\ref{Hamiltonc}) and (\ref{fMaxwell}).

To solve equations (\ref{Hamiltonc}) and (\ref{fMaxwell}), we start with a
stationary and axisymmetric metric ansatz describing a charged rotating
black hole localized on the 3-brane.
We write the metric in the Kerr-Schild form \cite{ks} with the
metric expressed in a linear approximation around the background:
\begin{eqnarray}
g_{\mu\nu}=\bar{g}_{\mu\nu}+v k_{\mu}k_{\nu}
\end{eqnarray}
where $v$ is an arbitrary scalar function and $k_{\mu}$ is null,
geodesic with respect to both the background metric $\bar{g}_{\mu\nu}$
and the full metric $g_{\mu\nu}$.

In general, it can be shown that if the vector field $k^{\mu}$
is null-geodesic, then Ricci tensor of $g_{\mu\nu}$ is related
to that of $\bar{g}_{\mu\nu}$ by \cite{dg:plb}
\begin{eqnarray}
R^{\mu}_{~\nu}=\bar{R}^{\mu}_{~\nu}-s^{\mu}_{~\rho}\bar{R}^{\rho}_{~\nu}
+\frac{1}{2}\bar{D}_{\rho}\bar{D}_{\nu} s^{\mu\rho}+\frac{1}{2}\bar{D}^{\rho}\bar{D}^{\mu} s_{\nu\rho}
-\frac{1}{2}\bar{D}^{\rho}\bar{D}_{\rho} s^{\mu}_{~\nu},
\end{eqnarray}
where $s_{\mu\nu}=v k_{\mu}k_{\nu}$ and $\bar{D}_{\mu}$ is the covariant
derivative with respect to the background metric $\bar{g}_{\mu\nu}$.
Thus, in our case if $\bar{g}_{\mu\nu}$ satisfies $\bar{R}_{\mu\nu}=\Lambda \bar{g}_{\mu\nu}$,
then the full metric $g_{\mu\nu}$ satisfies the Einstein
equations $R_{\mu\nu}=\Lambda g_{\mu\nu}+\kappa_{4}^2 T_{\mu\nu}$ with the same cosmological
constant, provided that $s_{\mu\nu}$ satisfies the linearized Einstein equations
with respect to the background metric $\bar{g}_{\mu\nu}$.

In this paper we will mainly deal with  spacetime(4D) with
constant scalar curvature.
In the previous section we saw that this choice leaves us two possibilities, flat or de-Sitter spacetime.
 In the flat ($R=0$) case, the solution for equations (\ref{Hamiltonc}) and (\ref{fMaxwell}) is given by
 the usual Kerr-Newman solution in general relativity \cite{ley}.
Thus from now on we concentrate ourselves on the de-Sitter case,
and choose the de-Sitter background metric in the Kerr-Schild form to solve
equations (\ref{Hamiltonc}) and (\ref{fMaxwell}).

Introducing the coordinates $x^{\mu}=\{\tau,r,\theta,
\varphi\}$, we write the metric in the Kerr-Schild form as
\cite{glpp}
\begin{eqnarray}
\label{metricsol}
ds^2&=&\left[-(1-\lambda r^2)\frac{\Delta_{\theta}}{\Xi}d\tau^2+
\frac{\Sigma}{(1-\lambda r^2)(r^2+a^2)}dr^2+
\frac{\Sigma}{\Delta_{\theta}}d\theta^2 +\frac{(r^2+a^2)\sin^2\theta}{\Xi}d\varphi^2\right] \nonumber \\
&& +H(r,\theta)\left[\frac{\Delta_{\theta}}{\Xi} d\tau
+\frac{\Sigma}{(1-\lambda r^2)(r^2+a^2)} dr
-\frac{a\sin^2\theta}{\Xi} d\varphi\right]^2,
\end{eqnarray}
where
\begin{eqnarray}
\Sigma &=& r^2+a^2\cos^2\theta, \nonumber \\
\Delta_{\theta} &=& 1+\lambda a^2\cos^2\theta, \nonumber \\
\Xi &=& 1+\lambda a^2.
\end{eqnarray}
Here $a$ is related to the angular momentum of the black
hole and $\lambda$ is a constant parameter determining the curvature of the background metric.

With the de-Sitter metric (\ref{metricsol}), the Maxwell equation (\ref{fMaxwell}) is satisfied if the potential
one-form is given by
\begin{equation}
\label{potential}
A_{\mu}dx^{\mu}=\frac{Qr}{\Sigma\Xi}
(-\Delta_{\theta}d\tau+ a\sin^2\theta d\varphi),
\end{equation}
where the parameter $Q$ is the electric charge of the black hole.
The nonvanishing components of the electromagnetic field tensor
$F_{\mu\nu}$ for the above potential one-form $A$ is calculated to be
\begin{eqnarray}
\label{em}
F_{r\tau}\!\!\!&=&\!\!\! \frac{Q\Delta_{\theta}(r^2-a^2\cos^2\theta)}{\Xi\Sigma^2},~
F_{\theta\tau}=-\frac{(1-\lambda r^2)Qra\sin2\theta}{\Xi\Sigma^2},
\nonumber \\
F_{r\varphi}\!\!\!&=&\!\!\! -\frac{(r^2-a^2\cos^2\theta)Qa\sin^2\theta}{\Xi\Sigma^2},~
F_{\theta\varphi}=\frac{(r^2+a^2)Qra\sin2\theta}{\Xi\Sigma^2}.
\end{eqnarray}
Using (\ref{em}) and setting the scalar function $H$ in the metric ansatz (\ref{metricsol}) as
\begin{eqnarray}
\label{HandP}
H(r,\theta)=r P(r,\theta)/\Sigma,
\end{eqnarray}
the Hamiltonian constraint (\ref{Hamiltonc})
 can be expressed as
\begin{eqnarray}
\label{Hamiltoncform}
0\!\!\!&=&\!\!\! -72\lambda(\lambda-m_c^2)\Sigma^2+6(m_{c}^2-2\lambda) \Sigma (2Z+rZ_{r})+r^2Z_{r}^2 \nonumber \\
&&\!\!\! -\frac{8r}{\Sigma}\left(r^2-\frac{a^2 \cos^{2}\theta}{2}\right)Z Z_{r}
+\frac{4}{\Sigma^2}(r^4-4a^2r^2\cos^{2}\theta+a^4\cos^{4}\theta)Z^2  \nonumber \\
&&\!\!\!-\frac{12Q^2}{\Sigma^{2}}[2(r^2-a^2\cos^{2} \theta)Z-r\Sigma Z_{r}]+\frac{24Q^4}{\Sigma^2},
\end{eqnarray}
where $Z=\partial P/\partial r$ and
$Z_{r}=\partial^2 P/\partial r^2$.
For $\lambda=0$ and $a=0$, the above equation reduces to the equation for
charged black holes in a flat background \cite{el}.

In order to find a general solution we solve (\ref{Hamiltoncform}) for $Z_{r}$:
\begin{eqnarray}
\label{Zr}
-rZ_{r}\!\!\!&=&\!\!\! \frac{6Q^2}{r^2(1+\alpha)}
-\left(\frac{4+2\alpha-2\alpha^2}{1+2\alpha+\alpha^2}\right)Z
+3(m_c^2 -2\lambda)r^2(1+\alpha) \nonumber \\
&&\!\!\! \pm 3m_c^2 r^2(1+\alpha)\sqrt{1+4u+\frac{4}{3}u^2},
\end{eqnarray}
where $\alpha(r,\theta)=a^2\cos^2\theta/r^2$ and $u=[Q^2-r^2(Z+3\lambda r^2(1+\alpha)^2)]/m_c^2 r^4(1+\alpha)^2.$
Introducing a new function $U$ by putting $Z$ as
\begin{equation}
\label{randZ}  Z=Q^2/r^2-3m_c^2 r^2(1+\alpha)^2U(z,\theta)/2
-3\lambda r^2(1+\alpha)^2,
\end{equation}
where $r=r_{0}e^{z}$, we express (\ref{Hamiltoncform}) in terms of
$U$:
\begin{eqnarray}
\label{lr}
&&\left[ \frac{2Q^{2}}{r^2}+\frac{3m_{c}^{2}r^{2}}{2}
 (2(1-\alpha^2)U+(1+\alpha)^2U_{z})+6\lambda r^2(1-\alpha^2) \right] \nonumber \\
&&=
\left[ \frac{2Q^{2}}{r^2}+\frac{3}{2}m_c^2 r^2(4+2\alpha-2\alpha^2)U
+6\lambda r^2(1-\alpha^2)+3m_c^2 r^2 (1+\alpha)(1\pm f)
\right],
\end{eqnarray}
where $f=\sqrt{1+6U+3U^2}$.
Solving (\ref{lr}) in terms of $U_{z}=\partial U/\partial z$, we get
\begin{equation}
\label{Ueq}
U_{z}=\frac{2}{1+\alpha}(1+U \pm f).
\end{equation}
Here we would like to mention two special solutions, $U=0$ and
$U=-2$ for minus and plus signs, respectively, for the last term in
the above equation.
Integrating (\ref{Ueq}) using the relation $r=r_{0}e^{z}$, the
solutions for $U \neq 0$  can be given implicitly by
\begin{eqnarray}
\label{regulU}
k_1\Sigma^4= -\frac{(1 + 3U + f)}{U^2 (3 + 3U +
       \sqrt{3}f)^{2\sqrt{3}}(- 5 - 3U + f)}
\end{eqnarray}
for the minus($-$) sign in the last term in (\ref{Ueq}), and for $U \neq -2$,
\begin{eqnarray}
\label{accelU}
  k_2\Sigma^4=- \frac{(- 5 - 3U + f)(-3 - 3U -
       \sqrt{3}f)^{2\sqrt{3}}}{(U + 2)^2( 1 + 3U + f)}
\end{eqnarray}
for the plus($+$) sign.
The above results reduce to the charged black hole case in
the limit $a \rightarrow 0$, and  $k_1$, $k_2$ are functions of the angle $\theta$
to be determined from appropriate boundary conditions.
The above two solutions correspond to the so-called regular and accelerated
branches \cite{gi:prd,gi:plb}.
Once $U$ is obtained, integrating (\ref{randZ}) using $Z=\partial P/\partial r$
we get the expression for $P(r,\theta)$ given below.
Then the scalar function $H(r,\theta)$ for the metric is given via (\ref{HandP}).
\begin{eqnarray}
\label{P}
 P(r,\theta)=-\frac{Q^2}{r}-\frac{\lambda}{r}\left(r^4+6a^2 r^2\cos^2\theta-3a^4\cos^4\theta\right)
 - \frac{3}{2}m_{c}^{2} \int dr ~r^{2}(1+\alpha)^2 U(r,\theta)
\end{eqnarray}
 For the two special solutions, $U=0$ and $U=-2$,
 they correspond to $\lambda=0$ and $\lambda=m_c^2$ in (\ref{P})
 and have Ricci scalar values $R=0$ and $R=12 m_c^2$, respectively.
 This coincides with the values we obtained in the previous section
 for spacetime with constant scalar curvature.
\\




Next, in order to check the physical properties of the metric given by
(\ref{metricsol}) we want to transform the
Kerr-Schild form to the Boyer-Lindquist coordinates.
For the case with constant curvature, the general Boyer-Lindquist type transformation is given by \cite{glpp}
\begin{eqnarray}
\label{toBL}
d\tau\!\!\!&=&\!\!\! dt+\frac{rP}{(1-\lambda r^2)\Delta_r}dr \nonumber \\
d\varphi \!\!\!&=& \!\!\!d\phi-\lambda a dt+\frac{a r P}{(r^2+a^2)\Delta_r}dr,
\end{eqnarray}
where $\Delta_r(r,\theta)=(r^2+a^2)(1-\lambda r^2)-rP(r,\theta)$.
The Maxwell equation (\ref{fMaxwell}) remains satisfied in the transformed Boyer-Lindquist coordinates,
if the potential one-form is given as follows.
\begin{equation}
\label{potential}
A_{\mu}dx^{\mu}=\frac{Q r}{\Sigma}\left(-dt
 +\frac{a\sin^2\theta}{\Xi}d\phi\right)
\end{equation}
Under the transformation (\ref{toBL}), the metric (\ref{metricsol}) takes
the form
\begin{eqnarray}
\label{BLT}
ds^2 = -\frac{\Delta_r}{\Sigma}\left(dt-\frac{a\sin^2\theta}{\Xi}d\phi\right)^2
+\frac{\Sigma}{\Delta_r}dr^2+\frac{\Sigma}{\Delta_{\theta}}d\theta^2
+\frac{\Delta_{\theta}\sin^2\theta}{\Sigma}\left[a dt-\frac{(r^2+a^2)}{\Xi}d\phi
\right]^2.
\end{eqnarray}
Now the constraint equation (\ref{Hamiltoncform}) for the
function $P(r,\theta)$ is
not preserved with the transformed metric (\ref{BLT}) if
$P$ has the $\theta$-angle dependence.
Since we want to keep the solutions obtained in the Kerr-Schild form
valid even in the transformed Boyer-Lindquist coordinates, we want to restrict
ourselves to the cases of $\theta$-independent $P$.
 From (\ref{P}), we see that  $P$ becomes independent of $\theta$
only in the following two cases, $\lambda=0,U=0$ and
$\lambda=m_c^2,U=-2$, and the both give the same scalar function $H$,
\begin{eqnarray}
\label{uniqueH}
H(r,\theta)=\frac{2Mr-Q^2}{\Sigma},
\end{eqnarray}
where $M$ is the total mass of the black hole, and boundary conditions are assumed
to be implemented.

Here, we note that the metric (\ref{BLT}) with the scalar function (\ref{uniqueH}) and $\lambda=m_c^2$
yields an exact Kerr-Newman-de Sitter solution with the cosmological constant $\Lambda=3m_c^2$,
where the crossover scale $r_c=m_c^{-1}$ behaves as an effective cosmological radius.

The self-accelerated branch of the Schwarzschild solution in the DGP model \cite{gi:prd,gi:plb} exhibits
the negative mass behavior, as its asymptotic behavior $P/r\approx -\tilde{r}_{M}^2/r^2 +m_c^2 r^2$ at the scale $r \gg r_*$ shows.
The term $-\tilde{r}_{M}^2/r^2$ looks like a 5D negative mass in the 5D viewpoint,
which suggests that there may be some non-perturbative instabilities in this branch.
In our constant curvature de-Sitter solution, the metric solution does not show such negative mass behavior at any scale
as we see below.
The Newton potential $\Phi$ which corresponds to  $-P/2r$ in the Schwarzschild case above can be
obtained from the metric solution (\ref{BLT})  as
\begin{equation}
-2\Phi=\frac{1}{r^2 + a^2 \cos^2 \theta} (2Mr-Q^2)+ m_c^2(r^2+a^2\sin^2\theta).
\end{equation}
The first term is the usual positive mass contribution to the potential, the second term is
 the usual charge contribution, and no negative mass contribution appears in the above potential.

Finally, in the de-Sitter solution the governing equation for horizon radius is given by
\begin{equation}
\label{horizon:con}
\Delta_r=(r^2+a^2)(1-m_c^2 r^2)-2Mr+Q^2=0.
\end{equation}
 and $r_{CH}$,
provided that the total mass $M$ lies in the range $M_{1e} \leq M \leq M_{2e}$
where $M_{1e}$ and $M_{2e}$ are given by \cite{dehghani:prd}
\begin{eqnarray}
\label{extrememass}
&M_{1e}&\!\!\!= \frac{1}{3\sqrt{6} m_c}
\sqrt{1+m_c^2(33a^2+36Q^2)(1-m_c^2a^2)-(1-14m_c^2a^2-12m_c^2Q^2+m_c^4a^4
)^{3/2}}, \nonumber \\
&M_{2e}&\!\!\!= \frac{1}{3\sqrt{6} m_c}
\sqrt{1+m_c^2(33a^2+36Q^2)(1-m_c^2a^2)+(1-14m_c^2a^2-12m_c^2Q^2+m_c^4a^4
)^{3/2}}. \nonumber \\
\end{eqnarray}
Here $r_{CH}$, which is smaller than the crossover scale $r_{c}$,
is a cosmological horizon, and $r_{+}$, $r_{-}$ are outer and inner
horizons, respectively.
Note that the horizons $r_{\pm}$ and $r_{CH}$ always have real
positive values if the total mass lies between the masses
$M_{1e}$ and $M_{2e}$.
\\


\section{Cosmological solution with de-Sitter background}

In this section, we will explicitly show that in the non-rotating limit our de-Sitter solution matches with a
 cosmological solution with accelerated expansion.
In showing this, we follow the method of matching conditions used in Ref. \cite{lss}.
We shall see that the solution with $\lambda=m_c^2$ as the accelerating universe with
the Hubble parameter $\mathcal{H}=r_c^{-1}$ in the non-rotating limit.
In \cite{lss}, requiring that the underlying gravity theory respect Birkhoff's
law  the modified gravitational force law needed
to generate the given cosmology was derived.
Here we reverse the process: We examine
the cosmological evolution corresponding to our metric (\ref{uniqueH}) with $\lambda=m_c^2$
assuming the Birkhoff's law in the non-rotating limit.

 Following the approach of Ref. \cite{lss} we consider a uniform sphere of dust and radiation(charge)
 and imagine that the evolution inside the sphere is exactly cosmological,
 while outside the sphere is filled with nonzero electric field.
 The mass and charge as the source are assumed to be unchanged throughout its time evolution.
 We take the cosmological metric to be the Robertson-Walker metric(flat space) inside the sphere:
\begin{eqnarray}
\label{rw}
ds^2 \!\!\!&=&\!\!\!-dt^2+a^2(t)\delta_{ij}dx^i dx^j \nonumber \\
     \!\!\!&=& \!\!\!-dt^2+a^2(t)[d\eta^2+\eta^2 d\Omega^2],
\end{eqnarray}
where $d\Omega^2$ is two-sphere.
Outside the sphere we take a Schwarzschild-like metric
\begin{eqnarray}
\label{outsch}
ds^2=-g_{00}(r)dT^2+g_{rr}(r)dr^2+r^2 d\Omega^2,
\end{eqnarray}
and rewrite it in a new form
\begin{eqnarray}
\label{newform}
ds^2=-N^2(t,\eta)dt^2+ a^2(t)d\eta^2+r^2 d\Omega^2,
\end{eqnarray}
where $r=r(t,\eta)$.
Now
we set the coordinate transformation between (\ref{outsch}) and (\ref{newform}) by
\begin{eqnarray}
\label{coordt}
dT=\dot{T} dt+T'd\eta,~~dr=\dot{r}dt+r'd\eta,
\end{eqnarray}
where the dot and prime denote partial differentiations with respect to t and $\eta$, respectively.
Rewrite (\ref{outsch}) under the transformation (\ref{coordt}) and compare it with (\ref{newform}),
 we obtain the following set of relations:
\begin{eqnarray}
\label{schnew}
N^2 \!\!\!&=&\!\!\! g_{00}\dot{T}^2-g_{rr}\dot{r}^2, \nonumber \\
0 \!\!\!&=&\!\!\! g_{00}\dot{T}T' -g_{rr}\dot{r} r', \nonumber \\
-a^2 \!\!\!&=& \!\!\!g_{00}T'^2-g_{rr} r'^2.
\end{eqnarray}
In order to match the metric (\ref{rw}) inside the sphere with the metric (\ref{newform}) outside,
we take the bounding surface
(boundary) of the sphere to be fixed at $\eta=\eta_{*}$.
By requiring continuity and smoothness at the boundary,
we obtain the following matching boundary conditions:
\begin{eqnarray}
\label{boundary}
r(t,\eta_*) \!\!\!&=&\!\!\! \eta_* a(t), \nonumber \\
r'(t,\eta_*) \!\!\!&=&\!\!\! a(t), \nonumber \\
N(t,\eta_*) \!\!\!&=&\!\!\! 1, \nonumber \\
N'(t,\eta_*) \!\!\!&=&\!\!\! 0.
\end{eqnarray}
Applying the above matching conditions to (\ref{schnew}), the
components of the metric in (\ref{outsch}) are determined in terms of $a(t)$:
\begin{eqnarray}
\label{incomp}
g_{00}=C^2 (1-\eta_{*}^2 \dot{a}^2 ),~~g_{rr}^{-1}=1-\eta_{*}^2 \dot{a}^2,
\end{eqnarray}
where $C^2$ is an integration constant.

In the non-rotating limit, from (\ref{BLT}) with (\ref{uniqueH}) for $\lambda=m_c^2$
the components of our metric outside the sphere
can be written  as
\begin{eqnarray}
\label{outcomp}
g_{00}=g_{rr}^{-1}= 1-\frac{2M}{r}-m_c^2 r^2+\frac{Q^2}{r^2}.
\end{eqnarray}
Comparing (\ref{incomp}) with (\ref{outcomp}) and setting $C^2=1$, we obtain a
Friedmann-like equation for the cosmological scale factor $a(t)$:
\begin{eqnarray}
\label{Friedmann}
\mathcal{H}^2=\left(\frac{\dot{a}}{a}\right)^2 =2M \eta_{*}^{-3} a^{-3}
+m_c^2-Q^2 \eta_{*}^{-4}a^{-4}.
\end{eqnarray}
When the scale factor $a(t)$ goes to infinity at large time,
the Hubble parameter $\mathcal{H}$ approaches the limit value
$\mathcal{H}_{0}=m_c=r_c^{-1}$ and yields the de Sitter expansion:
\begin{eqnarray}
\label{dSexp}
a(t) \propto e^{\mathcal{H}_{0}t}.
\end{eqnarray}
 Note that the cosmological metric (\ref{rw}) with (\ref{dSexp}) inside the sphere yields
 a constant Ricci scalar $12m_c^2$ at large time, which is the same value from our metric
(\ref{outcomp}).
\\

\section{Conclusion}


In this paper we investigate the Kerr-Newman solution on the braneworld with de-Sitter background
in the DGP model .
Assuming a $Z_2$-symmetry across the brane and
a stationary and axisymmetric metric ansatz in the Kerr-Schild
form with a constant curvature background on the brane,
we solve the constraint equation from (4+1)-dimensional gravity.

 When we restrict the spacetime with constant curvature, only
 flat and a de-Sitter cases are allowed by the constraint equation.
When we transform the metric obtained in the
Kerr-Schild form to the Boyer-Lindquist coordinates, the constraint equation in terms
of the metric scalar function is preserved (so is the solution) only
when the scalar curvature has two specific values, $0$ and $12m_c^2$.
This is in accord with the restriction from the constraint equation
since a de-Sitter type metric is used as the background metric
in the Kerr-Schild form.

Unlike the self-accelerated branch of the Schwarzschild solution in the DGP model,
our de-Sitter type solution does not have a negative mass behavior.
This is checked by the inspection of our metric solution.
About the effect of the bulk on the brane solution,
we checked that there is no gravitational effect from the
extra dimension
in our constant curvature de-Sitter solution.
We show this by evaluating the projected Weyl tensor(on the brane)
which turns out to be only proportional to the energy-momentum tensor on the brane.

In conclusion, we find an exact solution of Kerr-Newman type in a de-Sitter background
in the Kerr-Schild form,
which is also well behaved under the Boyer-Lindquist transformation.
Using the method of matching boundary conditions in the non-rotating limit,
we show that this solution has the characteristic of cosmological solution with accelerated expansion.
\\


\section*{Acknowledgments}

The authors thank KIAS for hospitality during the time that this
work was done. D.L. was supported by the Astrophysical Research
Center of the Structure and Evolution of the Cosmos (ARCSEC),
which is supported by the Korea Science and Engineering Foundation
(KOSEF). \\


\end{document}